\documentclass[]{aa}
\usepackage{graphicx}
\usepackage{verbatim}
\begin{document}
   \title{The nuclei of comets 7P/Pons-Winnecke, 14P/Wolf and 92P/Sanguin\thanks{Based on observations collected at the European Southern Observatory, Chile. Proposal: ESO No. 72.C-0233.} 
   }
   \titlerunning{The nuclei of comets Pons-Winnecke, Wolf and Sanguin}
   
   \author{C. Snodgrass
          \inst{1}
          \and
          A. Fitzsimmons
	  \inst{1}
          \and
          S. C. Lowry
	  \inst{1}
          }

   \offprints{C. Snodgrass\\
              \email{C.Snodgrass@qub.ac.uk}}

   \institute{APS Division, School of Physics and Astronomy, The Queen's University of Belfast, Belfast BT7 1NN, UK
         }

   \date{Received <date> / Accepted <date>}

   \abstract{
   Jupiter Family comets (JFCs) are short period comets which have recently entered the inner solar system, having previously orbited in the Kuiper Belt since the formation of the planets. We used two nights on the 3.6m New Technology Telescope (NTT) at the European Southern Observatory, to obtain $VRI$ photometry of three JFCs; 7P/Pons-Winnecke, 14P/Wolf and 92P/Sanguin. These were observed to be stellar in appearance. We find mean effective radii of $2.24 \pm 0.02$ km for 7P, $3.16 \pm 0.01$ km for 14P and $2.08 \pm 0.01$ km for 92P, assuming a geometric albedo of 0.04. From light-curves for each comet we find rotation periods of $7.53 \pm 0.10$ and $6.22 \pm 0.05$ hours for 14P and 92P respectively. 7P exhibits brightness variations which imply a rotation period of $6.8 \le P_{\mathrm{rot}} \le 9.5$ hours. Assuming the nuclei to be ellipsoidal the measured brightness variations imply minimum axial ratios $a/b$ of $1.3 \pm 0.1$ for 7P and $1.7 \pm 0.1$ for both 14P and 92P. This in turn implies minimum densities of $0.23 \pm 0.08$ g~cm$^{-3}$ for 7P, $0.32 \pm 0.02$ g~cm$^{-3}$ for 14P and $0.49 \pm 0.06$ g~cm$^{-3}$ for 92P. Finally, we measure colour indices of $(V-R) = 0.40 \pm 0.05$ and $(R-I) = 0.41 \pm 0.06$ for 7P/Pons-Winnecke, $(V-R) = 0.57 \pm 0.07$ and $(R-I) = 0.51 \pm 0.06$ for 14P/Wolf, and $(V-R) = 0.54 \pm 0.04$ and $(R-I) = 0.54 \pm 0.04$ for 92P/Sanguin.
   \keywords{comets: general -- comets: individual: 7P/Pons-Winnecke -- comets: individual: 14P/Wolf -- comets: individual: 92P/Sanguin -- techniques: photometric
	       }
   }

   \maketitle
%

\section{Introduction}

Cometary nuclei are some of the most unaltered bodies in the solar system; below a thin crust they very likely possess ices that first condensed out of the pre-solar nebula at the time the planets began to form. By studying comets we are looking at remnants of the building blocks left over from the formation of the planets. Cometary nuclei are, however, very difficult to observe. They are typically small (radii of a few kilometers) and dark (albedos approximately 4\%), making them very faint targets. A larger difficulty arises from the activity of comets; when they are near to the Sun the volatile ices that make up the nucleus sublimate to produce a coma. The far greater cross-sectional area of the coma means that it dominates the reflected flux when the comet is active. To isolate the flux from the nucleus in an active comet requires a very high spatial resolution image, so that the central pixels are dominated by the nucleus and the nucleus and coma components can be modelled separately. This normally requires Hubble Space Telescope resolution and/or a very close pass of the comet to the Earth (Lamy et al. \cite{La1999}; Lamy et al. \cite{La2002}). 

Alternatively, observations of comets can be performed at large heliocentric distances, when the comets are more likely to be inactive. The most abundant species in cometary nuclei is water ice, whose sublimation rate greatly increases when the comet comes within approximately 3 AU of the Sun. Hence such observations normally require nuclei to be beyond 3 AU, which requires the use of large optical telescopes.

Previously we have performed a programme (Lowry et al. \cite{lo2003}; Lowry \& Weissman \cite{LW2003}; Lowry \& Fitzsimmons \cite{LF2005}) of snap-shot photometry of Jupiter Family comets (JFCs), which are short period comets with periods less than 20 years, whose orbits are strongly influenced by Jupiter. These objects have largely prograde, low inclination orbits, and are thought to have only recently entered the inner solar system from the Kuiper Belt (Ip \& Fernandez \cite{if1997}). Around a third of numbered JFCs have had their radii measured by snap-shot surveys; in addition to our own, there are large ground based surveys by Licandro et al. (\cite{li2000}) and Meech et al. (\cite{M2004}). These results and others have been compiled by Lamy et al. (\cite{lamy}), producing a list which allows us to select targets with reliable radii, and therefore predictable brightnesses, for further study. To constrain the shape and internal structure of nuclei a series of observations over a number of days is required. Time series photometry allows us to plot a light-curve of the changing brightness of a rotating, non-spherical nucleus against time, and from this find a rotation period and limits on the elongation and bulk density of the nucleus. This paper describes our photometric observations of three JFC nuclei.

In section \ref{obs} we describe the observations and data reduction procedures employed in this work, before we explain our analysis methods, and present our results for each comet, in section \ref{results}. Section \ref{discussion} discusses our findings in terms of results for other comets, and looks at what we can say about the population of JFCs as a whole. Finally, section \ref{summary} summarises all of our results.


\section{Observations and data reduction}\label{obs}

We selected targets with reasonably well known radii from the compilation of Lamy et al. (\cite{lamy}), which allowed us to calculate approximate expected apparent magnitudes at the time of observation. In this way we were able to select three comets which were at large enough heliocentric distance to be inactive, but still observable at reasonable $S/N$, with $R$-band magnitudes $\sim$ 22. Comet 7P/Pons-Winnecke (hereafter 7P/PW) was discovered in 1819, and has an orbital period of 6.4 years. It has been previously observed when inactive by Lowry \& Fitzsimmons (\cite{LF2001}), who obtained a radius of 2.6 km with a snap-shot observation and an assumed albedo of 0.04. Comet 14P/Wolf (14P/W) was discovered in September 1884, just before its closest encounter with the Earth. It has an orbital period of 8.2 years, and has a previous radius and colour determination of 2.3 km and $(V-R) = 0.02 \pm 0.22$ determined from snap-shot observations (Lowry et al. \cite{lo2003}). Comet 92P/Sanguin (92P/S) was discovered in 1977, and orbits the Sun with a period of 12.4 years. Snap-shot observations of 92P/S were obtained on two occasions, by Lowry \& Weissman (\cite{LW2003}) and by Meech et al. (\cite{M2004}), who found radii of 1.7 and 1.2 km respectively.

\begin{table*}
\begin{minipage}[t]{2\columnwidth}

\caption{Summary of observations obtained using NTT+EMMI.}             
\label{observations}      
\centering                         
\renewcommand{\footnoterule}{}  
\begin{tabular}{l c c c c c c}        
\hline\hline                 
Comet & UT Date & $R_\mathrm{h}$ [AU]\footnote{Superscripts $I$ and $O$ refer to whether the comet is inbound (pre-perihelion) or outbound (post-perihelion). The variations in $R_\mathrm{h}$, $\Delta$ and $\alpha$ over the course of any one night were smaller than the significance quoted here.} & $\Delta$ [AU] & $\alpha$ [deg.] & Filter & Exposure time [s]\\    
\hline                        
7P/Pons-Winnecke & 20/01/2004 & 4.69$^I$ & 4.30 & 11.6 & 28$\times$$R$, 2$\times$$V$, 2$\times$$I$ & 120\\      
 & 21/01/2004 & 4.69$^I$ & 4.32 & 11.6 & 14$\times$$R$, 2$\times$$V$, 2$\times$$I$ & 120\\      
14P/Wolf & 20/01/2004 & 5.51$^O$ & 4.96 & 8.9 & 29$\times$$R$, $V$, $I$ & 220\\      
 & 21/01/2004 & 5.51$^O$ & 4.95 & 8.8 & 29$\times$$R$, $V$, $I$ & 220\\      
92P/Sanguin & 20/01/2004 & 4.46$^O$ & 3.58 & 6.3 & 63$\times$$R$, 3$\times$$V$, 3$\times$$I$ & 75\\      
 & 21/01/2004 & 4.46$^O$ & 3.59 & 6.4 & 49$\times$$R$, 3$\times$$V$, 3$\times$$I$ & 75\\      
\hline                                   
\end{tabular}
\end{minipage}
\end{table*}

Observations of the three comets were taken using the 3.6m New Technology Telescope (NTT) at the European Southern Observatory's (ESO) La Silla site, on the nights of 20th and 21st January 2004. At this time the comets were at heliocentric distances of $4.5 \le R_\mathrm{h} \le 5.5$ AU, and near opposition (Table \ref{observations}). CCD imaging was performed using the red arm of the EMMI instrument which is mounted at the f/11 Naysmith-B focus of the NTT. The EMMI red arm contains a mosaic of 2 MIT/LL 2048$\times$4096 CCDs, and was used in 2$\times$2 binning mode to give a pixel scale of 0.332 arcsec per pixel. The effective field of view is 9.1$\times$9.9 arcmin$^2$. The images were taken primarily through the Bessel $R$ band filter, with one set of images taken through the Bessel $V$ and $I$ filters each night for each comet to allow measurement of colour indices. All images were taken with the telescope tracking at the sidereal rate, with exposure times chosen so that the apparent motion of the comet would be less than 0.5\arcsec{}, and would thus remain within the seeing disk. This allowed us to perform accurate measurement of the stellar-background point spread function (PSF), and highly accurate differential photometry. The $FWHM$ of the stellar-background PSF was measured to be vary between 0.6\arcsec{} and 1.5\arcsec{} over the two nights, with a median of 0.8\arcsec. The $VRI$-filter exposure times used were 120 seconds for comet 7P/PW, 220 seconds for 14P/W and 75 seconds for 92P/S. We observed each comet in blocks of exposures lasting $\sim$ 20 minutes, and cycled between each of the comets that were visible at any given point in the night. This procedure gave us good temporal coverage without losing too much time slewing the telescope between targets.

The EMMI red arm camera reads out through four individual amplifiers (two per CCD chip) to improve read out times. Each readout is stored as a separate image within a multi-extension fits-format file. By careful positioning of the telescope, the comets and nearby comparison stars were kept within the same half of chip 1 at all times, lessening the reduction workload as only one fits extension needed be processed. The reduction was performed using standard IRAF tasks (Tody \cite{IRAF1}; Tody \cite{IRAF2}). Bias-subtraction was performed utilising the prescan strip of chip 1. Twilight sky exposures were acquired during the evening and morning, and these were combined to produce a flat-field image for each filter for each night. Comparison of evening and morning flats revealed a slight change in illumination of the focal plane, giving a variation of $\sim 1 \%$ in the overall field illumination from one side of the image to the other. However this is a known effect caused by the Naysmith mirror baffle (Hainaut \cite{NTT-illum}), and was correctable by averaging both sets of flats for each filter.

Differential photometry was performed on the comets using the IRAF packages DIGIPHOT and APPHOT (Davis \cite{IRAF3}). Aperture radii were set equal (to the nearest integer pixel) to the $FWHM$ of the stellar PSF, which has previously been found to be the optimum for maximising $S/N$ (Howell \cite{ap-FWHM}). Choosing aperture radii as a function of stellar PSF also counters the problem of variations in measured brightness with changing seeing, which is encountered when using small fixed apertures (Licandro et al. \cite{Lic2000}). In addition, photometry of the field stars used to measure the differential magnitudes, and of standard stars from the Landolt (\cite{landolt}) catalogue, was performed using an aperture of diameter 10\arcsec. The smaller aperture was used for differential photometry between the comets and field stars, while the larger diameter was used for photometric calibration and was that used by Landolt (\cite{landolt}). Using the IRAF package PHOTCAL these standard star measurements gave the zero-point, extinction coefficient and colour term for each filter for each night. Unfortunately, many of the standard star observations taken on the second night in the $R$ and $I$ bands were saturated. We calculated the extinction in the $R$-band from observations of field stars observed throughout the night; taking this as an input allowed us to produce satisfactory solutions for the zero-point and colour term using the non-saturated standard stars. For the $I$-band, in which all observations were taken at approximately constant airmass, we were forced to assume a constant zero-point, taking the values from the first night. As there was no change in instrumental set up and both nights were photometric, and the calculated zero-points in the $V$ and $R$-bands were found to be consistent across both nights, this assumption is acceptable. The photometric calibrations were then used to calculate the magnitudes of the field stars in each frame, taking the mean of these values gave us a very accurate measurement of the brightness of our comparison stars. Adding this value to each of the differential comet magnitudes gave us accurate calibrated comet magnitudes. Care was taken to select comparison field stars and standard stars with colours similar to the Sun, as the comets were expected to have near-solar colours. 

\section{Results}\label{results}

\subsection{7P/Pons-Winnecke}
\label{7Presults}

   \begin{figure}
   \centering
   \includegraphics[width=0.47\textwidth]{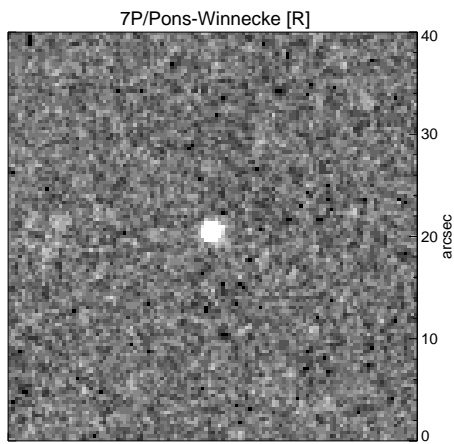}
      \caption{Co-added image of 7P/PW from 28 frames taken on the January 20th. Each frame had an exposure time of 120s, giving this combined frame an equivalent exposure time of 56 min. The shift-and-add procedure used to produce this image was designed in such a way to remove cosmic rays, field stars and the background sky, leaving only the comet.
              }
         \label{7Pimage}
   \end{figure}
%

   \begin{figure}
   \centering
   \includegraphics[angle=-90,width=0.47\textwidth]{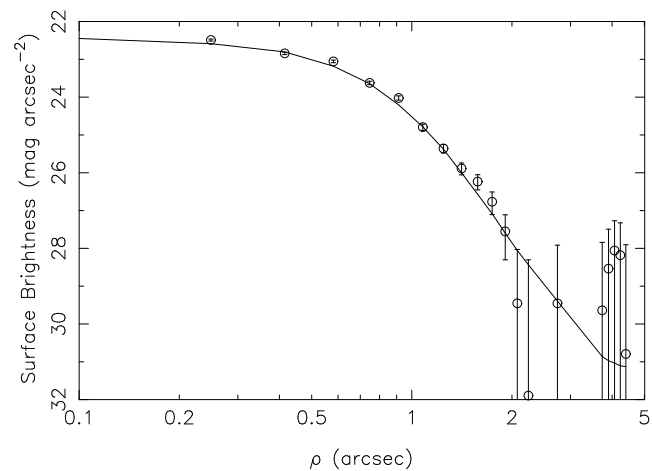}
      \caption{Surface brightness profile of 7P/PW. Here the surface brightness in mag.~arcsec$^{-2}$ is plotted as a function of radius $\rho$ from the centre of the comet in arcsec. The profile is indistinguishable from the scaled stellar PSF (the solid line), implying that the comet is a point-source, and therefore inactive. At a surface brightness of $\Sigma \approx 30$ the flux is only a few counts per arcsec$^2$.
              }
         \label{7Psbp}
   \end{figure}
%

7P/PW appears stellar in each of the 42 individual images. A combined image (Fig. \ref{7Pimage}) was produced by first combining all $R$-band images to calculate the median image for the data, to produce a deep image of the background star field without the comet or cosmic rays. A scaled version of this image was then subtracted from each individual $R$-band frame, removing fixed objects and the background sky. Each subtracted frame was then shifted to centre on the comet, and these were combined using a median filter to remove cosmic rays and leave only the comet. The image of 7P/PW produced using the $28 \times 120$ second $R$-band exposures taken on the first night is equivalent to a single 56 minute exposure. The brightness profile of the comet measured in this combined image is shown in Fig. \ref{7Psbp}. It is indistinguishable from the image PSF obtained from the profile of a bright star measured in the deep star field image. 

To put limits on any unresolved coma, we measure the surface brightness at a large distance $\rho$ (arcsec) from the centre of the comet image. We measure the surface brightness of the coma at $\rho$ = 5\arcsec{} to be $\Sigma_c(5) \ge$ 30.8 mag.~arcsec$^{-2}$. We set $\rho$ = 5\arcsec{} as at this radius we expect any flux to be dominated by any unresolved coma present. The measured $FWHM$ of the combined comet image is 1.1\arcsec{}. We can compare the integrated coma magnitude $m_c$ measured within this aperture with the calibrated total magnitude $m_R$, as we used an aperture of $\rho$ = 5\arcsec{} to calibrate our field stars, therefore including an implicit aperture correction in the comet magnitude. For a steady state coma, the surface brightness is inversely proportional $\rho$, and $m_c$ is given by (Jewitt \& Danielson \cite{JD84})
\begin{equation}\label{sbp-eqn}
m_c(\rho) = \Sigma_c(\rho) - 2.5\log(2\pi\rho^2).
\end{equation}
We find $m_c(5) \ge 25.3$. The integrated magnitude of the comet was found to be $m_R = 22.465 \pm 0.016$, meaning that any unresolved coma contributes $10^{0.4(m_R - m_c)} \le 7.3 \pm 16.0$\% of the flux from 7P/PW. We therefore believe the nucleus was effectively inactive during our observations.

The observed magnitude of a cometary nucleus is given by 
\begin{equation}
m_R = m_R(1,1,0) + 5\log(R_h \Delta) + \beta\alpha,
\end{equation}
where $m_R(1,1,0)$ is the `absolute' magnitude of the nucleus as it would be measured at a hypothetical point at heliocentric ($R_h$) and geocentric ($\Delta$) distances of 1 AU, and at phase angle $\alpha$ = 0\degr. As we only have observations over two nights at approximately constant $\alpha$, we cannot independently measure the phase coefficient $\beta$. We therefore take the commonly assumed value of $\beta$ = 0.035 mag.~deg$^{-1}$. For each of our comets, the variation in the last two terms was found to be negligible over the two days of observations; for 7P/PW the difference in $m_R$ due to these terms between the first and last frame was found to be $\delta m_R$ = 0.008 mag, considerably smaller than the uncertainty on the individual measurements. As the conversion to absolute magnitudes is therefore unnecessary, we choose to use apparent magnitudes to produce a light curve. The measured apparent $R$-band magnitudes for 7P/PW are listed in Table \ref{7Ptable}.

   \begin{figure}
   \centering
   \includegraphics[angle=-90,width=0.47\textwidth]{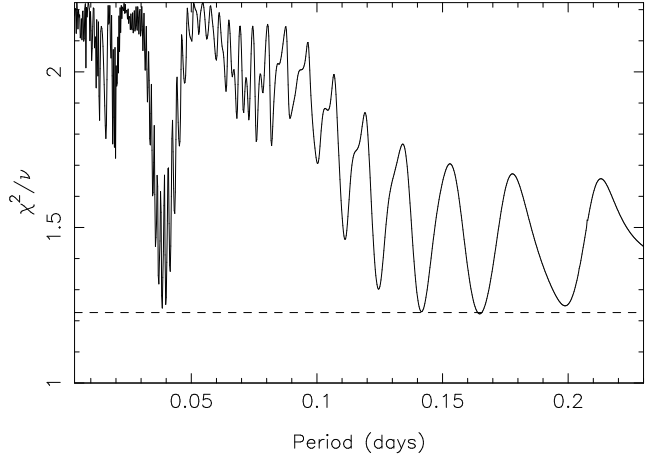}
      \caption{Periodogram for 7P/PW, showing reduced $\chi^2$ against period. The horizontal line shows the expected value of $\chi^2/\nu$ for $1\sigma$ residuals. There are a number of periods which give minima close to this line. The first can be rejected, and is due to the sampling period, but the group around $P_{\mathrm{fitted}} \sim$ 0.16 days all produce acceptable light-curves. The data do not allow us to choose one of these periods as the `correct' value, so we take $P_{\mathrm{rot}} = 7.9~{}^{+1.6}_{-1.1}$ hours, from the period with the deepest minima, but with the uncertainty encompassing the full range of acceptable periods.
              }
         \label{7Ppgram}
   \end{figure}
%

We use a version of the method of Lomb (\cite{Lomb}) to search for periodicities in the brightness variations of the comet. We fit a first-order Fourier model:
\begin{equation}
m_R(t) = C + A\cos{\omega t} + B\sin{\omega t}
\end{equation}
to the data for a range of frequencies $\omega$, finding the optimum coefficients $A$, $B$ and $C$ for each and calculating the reduced $\chi^2$ for each to produce a periodogram (Fig. \ref{7Ppgram}). The reduced $\chi^2$ is $\chi^2/\nu$, where $\nu$ is the number of degrees of freedom of the model, given by $\nu = (N - 3)$ where $N$ is the number of data points. For 7P/PW, $N$ = 42. A good fit to the data, where the scatter of residuals is a gaussian with $1\sigma$ variance, gives $\chi^2/\nu = 1 \pm \sqrt{2/\nu}$.

   \begin{figure}
   \centering
   \includegraphics[angle=-90,width=0.47\textwidth]{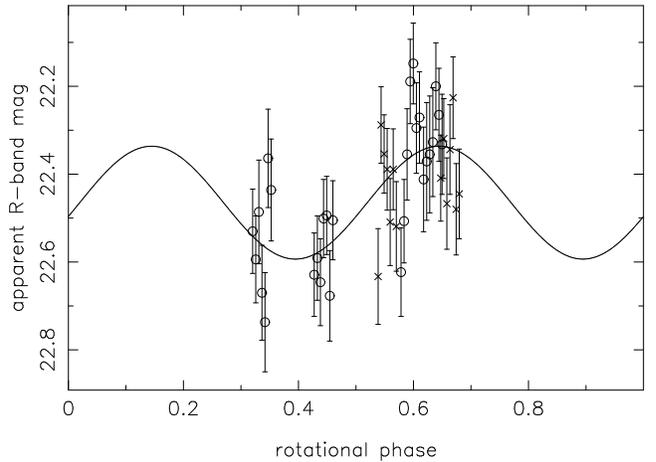}
      \caption{Folded light curve for 7P/PW, period = 7.89 hours. This was the strongest periodicity found, however the true period could lie anywhere in the region between 5.8 and 9.5 hours. We require further data to constrain the period. 
              }
         \label{7Pbest}
   \end{figure}
%

Figure \ref{7Ppgram} shows $\chi^2/\nu$ against period in days for 7P/PW. There are 4 minima which fall close to the horizontal line marking $\chi^2/\nu = 1 + \sqrt{2/\nu} = 1.23$. The first, at $\sim$ 1 hour, can be ignored as it is due to the data sampling period. The second grouping, with the strongest minimum at $P_{\mathrm{fitted}} = 0.165 \pm 0.006$ days $= 3.95 \pm 0.15$ hours, produce acceptable light-curves. With only two nights data, we cannot differentiate between the 3 minima; the central minimum is statistically acceptable with $\chi^2/\nu = 1.22 < 1.23$, but the others are close to this ($\chi^2/\nu = 1.23$ and 1.25) and all 3 produce equally acceptable light-curves, as determined by visual inspection of the phased data. We therefore quote our best-fit period to be $P_{\mathrm{fitted}} = 3.95 \pm {}^{0.80}_{0.54}$ hours, where the uncertainty covers the acceptable periods either side of the most probable period in Fig. \ref{7Ppgram}. Here we assume that the variation in brightness is due to the nucleus being a rotating non-spherical body, and that the rotation period $P_{\mathrm{rot}}$ is twice the fitted period $P_{\mathrm{fitted}}$. This gives $P_{\mathrm{rot}} = 7.9~{}^{+1.6}_{-1.1}$ hours and the expected double-peaked light-curve (Fig. \ref{7Pbest}).

Fitting a more general
\begin{equation}
m_R(t) = C + \sum_{l=1}^{M} A_l\cos{l\omega t} + B_l\sin{l\omega t}
\end{equation}
model to the data did not produce an improved $\chi^2/\nu$ for $M$ = 2 or higher orders, although with such sparse data it would have been difficult to justify a choice of a higher order fit without a significant improvement in reduced $\chi^2$. This more general model has $\nu = N - (2M + 1)$, taking into account the inclusion of more model parameters. For $M=1$ this reduces to the simple first order equation. Our null hypothesis is that there is no periodic variation; we fit a $M=0$ constant brightness model, which gives $\chi^2/\nu = 2.22 \approx 5.5\sigma$. We can therefore reject the null hypothesis at a 5$\sigma$ level and be confident that there is a real variation in the brightness of the comet.

The peak-to-trough range in brightness of the nucleus $\Delta m = 0.30 \pm 0.05$ mag. We assume that the brightness variations are due to the changing observed cross-section of a rotating elongated nucleus, as has been observed during comet fly-bys by spacecraft (Stooke \& Abergel \cite{sa1991}; Oberst et al. \cite{o2004}; Duxbury et al. \cite{d2004}). We model the nucleus as the simplest non-spherical shape, a tri-axial ellipsoid with semi-axes $a$, $b$ and $c$ where $a>b$ and $b=c$. The axial ratio $a/b$ is related to the range in observed magnitudes by
\begin{equation}\label{axial_ratio_eqn}
\frac{a}{b} \ge 10^{0.4\Delta m}.
\end{equation}
This gives a lower limit on $a/b$ as we measure only the axial ratio projected onto the plane of the sky, and we do not know the orientation of the rotation axis. This gives $a/b \ge 1.3 \pm 0.1$ for 7P/PW.

The mean measured apparent magnitude of comet 7P/PW was $m_R = 22.465 \pm 0.016$, which is related to the average radius of the nucleus $r_{\mathrm{N}}$ by
\begin{equation}\label{rneqn}
A_R r^2_{\mathrm{N}} = 2.238 \times 10^{22} R^2_h \Delta^2 10^{0.4(m_\odot - m_R + \beta\alpha)}
\end{equation}
where $A_R$ is the geometric albedo and $m_\odot=-27.09$ is the apparent magnitude of the Sun, both in the $R$-band (Russel \cite{R16}). The other terms are as defined above. The geometric albedo of cometary nuclei has previously been found to be very low (Lamy et al. \cite{lamy}); we take the commonly assumed value of $A_R = 0.04$ to give $r_{\mathrm{N}} = 2.24 \pm 0.02$ km. The uncertainties on this result, and on others given in this paper, are based purely on the uncertainty in the photometry, and do not take into account any uncertainties in the values of either $A_R$ or $\beta$. This mean radius gives us the size of the equivalent spherical nucleus; if we take into account our ellipsoidal model we obtain dimensions of the nucleus of $a \times b = 2.57 \times 1.95$ km. These results are consistent with the result found by Lowry \& Fitzsimmons (\cite{LF2001}), who measured a radius of $r_{\mathrm{N}} = 2.6 \pm 0.1$ km, if we assume that their snap-shot observation was taken at a light-curve maximum. 

We took one set of observations each night through $V$ and $I$ filters in addition to our time series photometry in the $R$-band. We interpolate between $R$-band measurements to find the apparent $m_R$ at the time of the $V$ and $I$-band observations, and thus measure the colour indices $(V-R)$ and $(R-I)$. For 7P/PW, we find $(V-R) = 0.40 \pm 0.05$ and $(R-I) = 0.41 \pm 0.06$, both of which are redder than the Sun and in the expected range for cometary nuclei (see section \ref{discussion}).


\subsection{14P/Wolf}

   \begin{figure}
   \centering
   \includegraphics[width=0.47\textwidth]{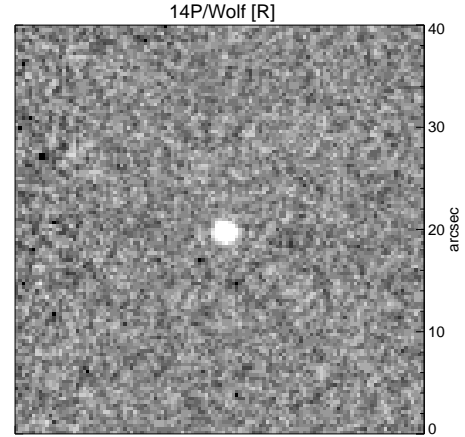}
      \caption{Co-added image of all 29 frames taken of 14P/W on the January 20th. Each frame had an exposure time of 220s, giving this combined frame an equivalent exposure time of 1.77 hr. 14P/W appears stellar; a surface brightness profile (Fig. \ref{14Psbp}) and calculated limits on the coma brightness show that it was most likely inactive at the time of observation.
              }
         \label{14Pimage}
   \end{figure}
%

   \begin{figure}
   \centering
   \includegraphics[angle=-90,width=0.47\textwidth]{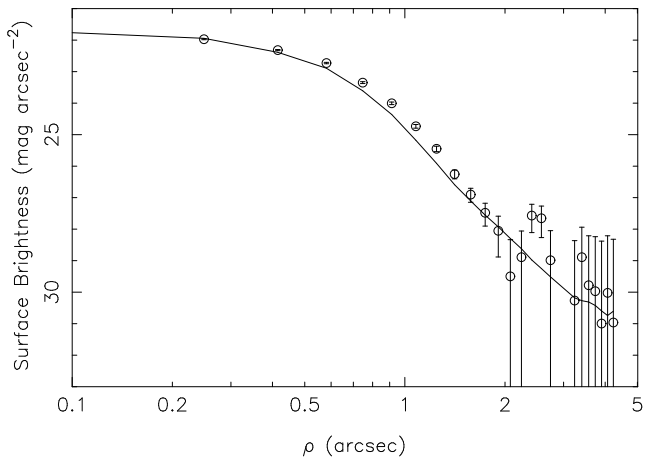}
      \caption{Same as Fig. \ref{7Psbp}, for 14P/W. The close match of the comet and stellar profiles, taken together with the calculated upper limit of 5\% on the flux from any coma, imply that the comet is inactive.
              }
         \label{14Psbp}
   \end{figure}
%

The same procedures were followed for 14P/W. Co-added $R$-band images of the 29 individual frames taken on each night appear stellar (Fig. \ref{14Pimage}), and measuring the surface brightness profile at large $\rho$ gave limits on the percentage of flux that could be from any unresolved coma (Fig. \ref{14Psbp}). For 14P/W the average coma surface brightness from the two nights was found to be $\Sigma_c(5) \ge$ 31.0 mag.~arcsec$^{-2}$. Using equation \ref{sbp-eqn} and the integrated magnitude of the comet $m_R = 22.281 \pm 0.007$, we obtain a limit on the flux from the coma of $\le 5.3 \pm 10.5\%$ of the observed flux of 14P/W. Again we therefore assume we observed an inactive nucleus.

   \begin{figure}
   \centering
   \includegraphics[angle=-90,width=0.47\textwidth]{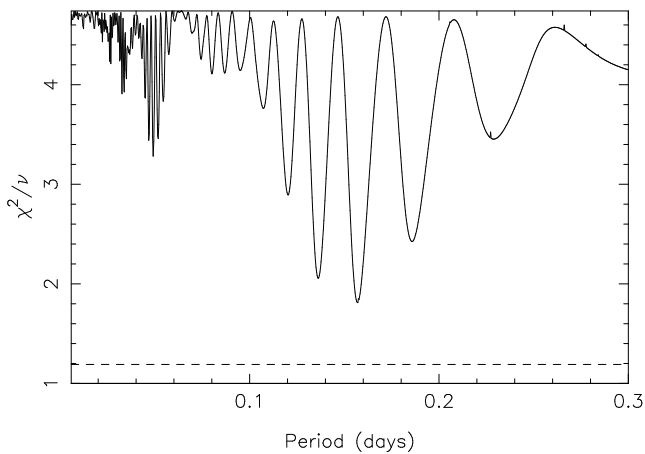}
      \caption{Same as Fig. \ref{7Ppgram}, for 14P/W. The strongest period found is at  $P_{\mathrm{fitted}} = 3.72 \pm 0.05$ hours, corresponding to rotation period of $P_{\mathrm{rot}} = 7.53 \pm 0.10$ hours.
              }
         \label{14Ppgram}
   \end{figure}
%

   \begin{figure}
   \centering
   \includegraphics[angle=-90,width=0.47\textwidth]{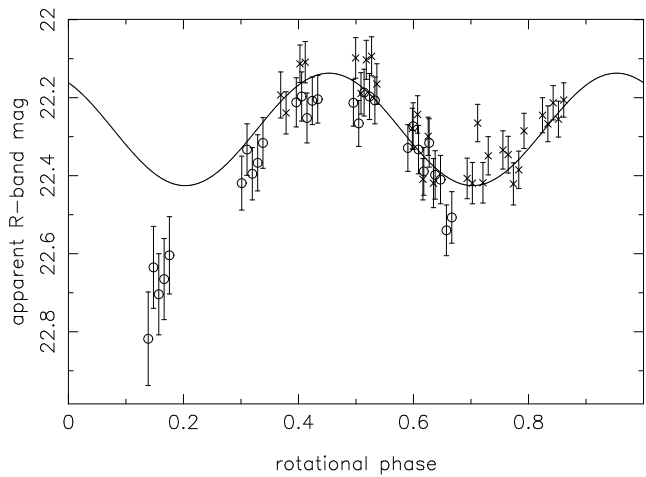}
      \caption{Folded light curve for 14P/W, period = 7.53 hours. 
              }
         \label{14Pbest}
   \end{figure}
%

Figure \ref{14Ppgram} shows the first order periodogram for 14P/W. Although there are no periods with the expected $\chi^2/\nu < 1.19$, there is a strong minimum at $P_{\mathrm{fitted}} \approx 0.16$ days which corresponds to a 4.3$\sigma$ fit. The null hypothesis, that the brightness of the comet is constant and the variations are due entirely to observational errors, is rejected at a 19$\sigma$ confidence level. There is clearly a periodic brightness variation, as can be seen in Fig. \ref{14Pbest}, which shows the photometric data folded onto twice the strongest period, assuming a double-peaked rotational light curve. This gives $P_{\mathrm{rot}} = 7.53 \pm 0.10$ hours. The shape of the light-curve in Fig. \ref{14Pbest} suggests that a higher order fit, with two minima of different depths, may give a better model to this data. This would suggest a departure from a simple tri-axial ellipsoid, as this sort of light-curve would be produced by a rotating body with, for example, a pear-like shape. The strongest minimum in a second order fit periodogram is at the same period, giving a slightly saw tooth shaped light-curve and a marginal reduction in $\chi^2/\nu$, but this still gives light-curve minima with equal depths. A group of minima around $P_{\mathrm{rot}} = 7.5$ hours do produce asymmetric light-curves, however they do not give a formal improvement on the fit as they do not lower $\chi^2/\nu$ relative to the 1st order fit. 

The observed data (Table \ref{14Ptable}) has a range in magnitudes $\Delta m = 0.55 \pm 0.05$ mag., which, using equation \ref{axial_ratio_eqn}, allows us to describe 14P/W as a tri-axial ellipsoid with axial ratio $a/b \ge 1.7 \pm 0.1$. However the axial ratio gives only a very simplified description of shape; it does not contain enough information to adequately describe an irregularly shaped nucleus. 

The median apparent magnitude of 14P/W was measured to be $m_R = 22.281 \pm 0.007$ mag., giving a mean effective radius of $r_{\mathrm{N}} = 3.16 \pm 0.01$ km, again assuming an albedo of 4\%. This implies dimensions of the nucleus of $a \times b = 4.07 \times 2.45$ km. This is slightly larger than the snap-shot radius found by Lowry et al. (\cite{lo2003}) of $r_{\mathrm{N}} = 2.33 \pm 0.12$ km, but consistent with the observed shape induced light-curve. 

We measured colour indices of $(V-R) = 0.57 \pm 0.07$ and $(R-I) = 0.51 \pm 0.06$ for 14P/W, placing it at the red end of the observed range in colours for JFCs. This is in sharp contrast with the colours measured by Lowry et al. (\cite{lo2003}), which at $(V-R) = 0.02 \pm 0.22$ and $(R-I) = 0.25 \pm 0.35$ made 14P/W the bluest cometary nucleus, and considerably bluer than the Sun. The large uncertainties are due to the poor atmospheric conditions experienced at the time of observation by this earlier work. Our new data provide a much more reliable measurement of the nucleus' surface colour, and we believe that the large change is due to improved measurement, not any physical change in the comet during the 4.5 years (including a perihelion passage in November 2000) between the observations.


\subsection{92P/Sanguin}

   \begin{figure}
   \centering
   \includegraphics[width=0.47\textwidth]{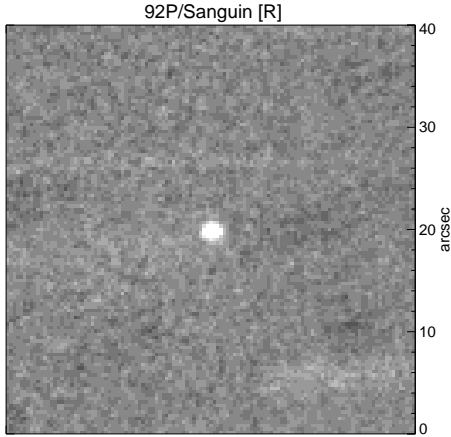}
      \caption{Co-added image of all 63 frames taken of 92P/S on the January 20th. Each frame had an exposure time of 75s, giving this combined frame an equivalent exposure time of 1.31 hr. 
              }
         \label{92Pimage}
   \end{figure}
%

   \begin{figure}
   \centering
   \includegraphics[angle=-90,width=0.47\textwidth]{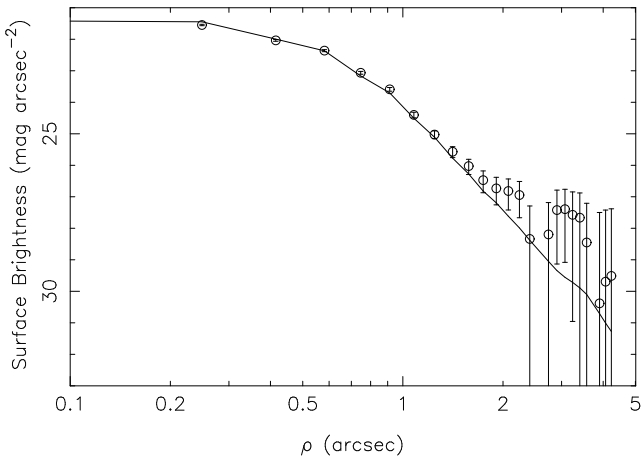}
      \caption{Same as Fig. \ref{7Psbp} for 92P/S. Again, the profile and 8\% coma upper limit imply that the comet was inactive at the time of observation.
              }
         \label{92Psbp}
   \end{figure}
%

92P/S appears stellar in both individual and co-added images (Fig. \ref{92Pimage}), and a surface brightness profile (Fig. \ref{92Psbp}) shows that it was inactive at the time of our observations. Again, we calculate limits on any unresolved coma by measuring $\Sigma_c(5) \ge$ 30.1 mag.~arcsec$^{-2}$. This corresponds to a coma magnitude of $m_c \ge 24.6$, equivalent to $\le 8.7 \pm 17.6\%$ of the average flux from the comet, which was found to have an integrated magnitude of $m_R = 21.938 \pm 0.007$.  

   \begin{figure}
   \centering
   \includegraphics[angle=-90,width=0.47\textwidth]{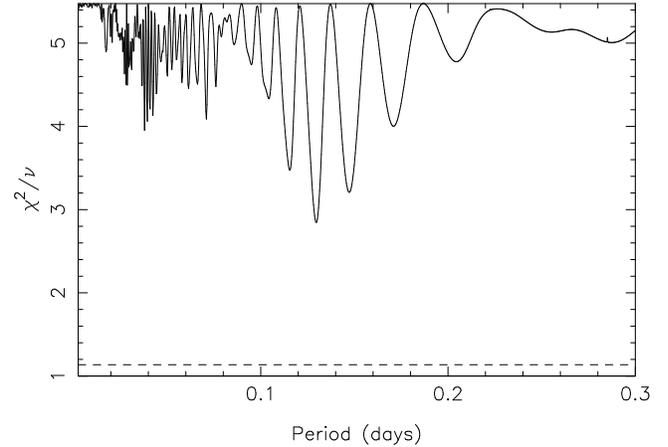}
      \caption{Same as Fig. \ref{7Ppgram} for 92P/S. The strongest period found is at $P_{\mathrm{fitted}} = 3.11 \pm 0.03$ hours, corresponding to rotation period of $P_{\mathrm{rot}} = 6.22 \pm 0.05$ hours.
              }
         \label{92Ppgram}
   \end{figure}
%

   \begin{figure}
   \centering
   \includegraphics[angle=-90,width=0.47\textwidth]{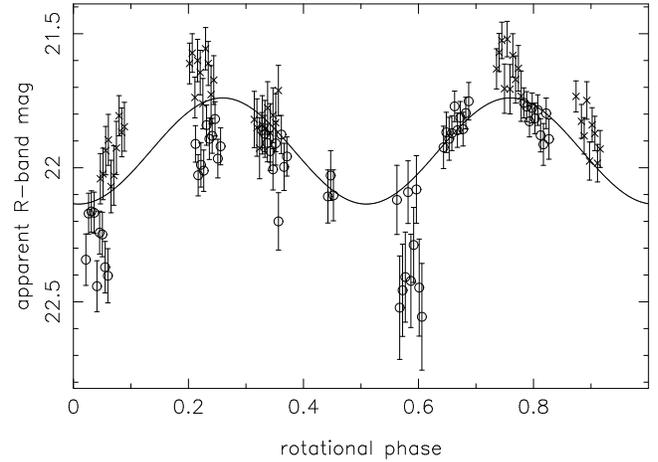}
      \caption{Folded light curve for 92P/S, period = 6.22 hours. 
              }
         \label{92Pbest}
   \end{figure}
%

Table \ref{92Ptable} gives the measured apparent $R$-band magnitudes for 92P/S. The periodogram for 92P/S (Fig. \ref{92Ppgram}) shows a clear minimum at 0.13 days, corresponding to a rotation period at $P_{\mathrm{rot}} = 2 \times P_{\mathrm{fitted}} = 6.22 \pm 0.05$ hours. This minima has $\chi^2/\nu = 2.85 \approx 14\sigma$. The null hypothesis (constant brightness) is rejected at a $39\sigma$ confidence level. The brightness variations are clearly real; the light-curve, folded onto the best-fit rotation period, is shown in Fig. \ref{92Pbest}. Second order or higher fits to the data do not formally improve upon the fit; the data which suggests to the eye a deeper second minimum, and therefore a higher order fit, may be an artefact of the larger scatter of points at lower $S/N$ at lower intrinsic brightness.

The data give $\Delta m = 0.60 \pm 0.05$ mag., corresponding to an axial ratio of $a/b \ge 1.7 \pm 0.1$. We measure a mean magnitude $m_R = 21.938 \pm 0.007$. Once again, we use equation \ref{rneqn}, assuming an albedo of 0.04, to give a radius of the equivalent spherical body of $r_{\mathrm{N}} = 2.08 \pm 0.01$ km. This gives the nucleus of 92P/S projected dimensions of $a \times b = 2.74 \times 1.57$ km. 

The radius of 92P/S has previously been measured to be $r_{\mathrm{N}} = 1.70 \pm 0.63$ km (Lowry \& Weissman \cite{LW2003}) and $r_{\mathrm{N}} = 1.18 \pm 0.20$ km from Meech et al. (\cite{M2004}), calculating $r_{\mathrm{N}}$ in this case from their result $m_R = 26.451 \pm 0.364$ using the position of the comet at the time of observation ($R_\mathrm{h}$ = 8.57 AU, $\Delta$ = 8.43 AU, $\alpha$ = 6\fdg6), $A_R = 0.04$ and $\beta = 0.035$ mag.~deg$^{-1}$ for consistency with our results. The first of these produces results consistent with our data, from a snap-shot taken at similar heliocentric distance ($R_\mathrm{h} \approx 4.5$ AU). The second result gives a lower radius, but is consistent at a 2$\sigma$ level with our measurements if we assume that this snap-shot was taken at a minimum in the rotational light-curve. Finally, we measured the colour indices of 92P/S to be $(V-R) = 0.54 \pm 0.04$ and $(R-I) = 0.54 \pm 0.04$.


\section{Discussion}\label{discussion}

We now look at these results in the context of other Jupiter Family comets, in order to attempt to draw some conclusions about the population as a whole. The sizes, spin rates and colours determined for these three comets are typical of the results determined for comets, all falling within previously observed ranges. However, adding three more comets to the database of known nuclear properties is a significant step towards having a large enough sample to enable statistically significant conclusions to be reached.

\setcounter{table}{4}

\begin{table*}
\begin{minipage}[t]{2\columnwidth}

\caption{Derived physical parameters and colours for 3 comets.}             
\label{results_table}      
\centering                         
\renewcommand{\footnoterule}{}  
\begin{tabular}{l c c c c c c c c c}        
\hline\hline                 
Comet & $m_R$ & $m_c$ & $m_R(1,1,0)$ & $r_{\mathrm{N}}$ [km] & $P_{\mathrm{rot}}$ [hr] & $a/b$ & $D_{\mathrm{N}}$ [g cm$^{-3}$] & $(V-R)$ & $(R-I)$\\    
\hline                        
7P/PW & 22.465$\pm$0.016 & $\ge$25.3 & 15.532$\pm$0.016 & 2.24$\pm$0.02 & 7.9${}^{+1.6}_{-1.1}$ & 1.3$\pm$0.1 & 0.22$\pm$0.08 & 0.40$\pm$0.05 & 0.41$\pm$0.06\\      
14P/W & 22.281$\pm$0.007 & $\ge$25.5 & 14.787$\pm$0.007 & 3.16$\pm$0.01 & 7.53$\pm$0.10 & 1.7$\pm$0.1 & 0.32$\pm$0.02 & 0.57$\pm$0.07 & 0.51$\pm$0.06\\      
92P/S & 21.938$\pm$0.007 & $\ge$24.6 & 15.700$\pm$0.007 & 2.08$\pm$0.01 & 6.22$\pm$0.05 & 1.7$\pm$0.1 & 0.49$\pm$0.06& 0.54$\pm$0.04 & 0.54$\pm$0.04\\   
\hline                                   
\end{tabular}
\end{minipage}
\end{table*}

   \begin{figure}
   \centering
   \includegraphics[angle=-90,width=0.47\textwidth]{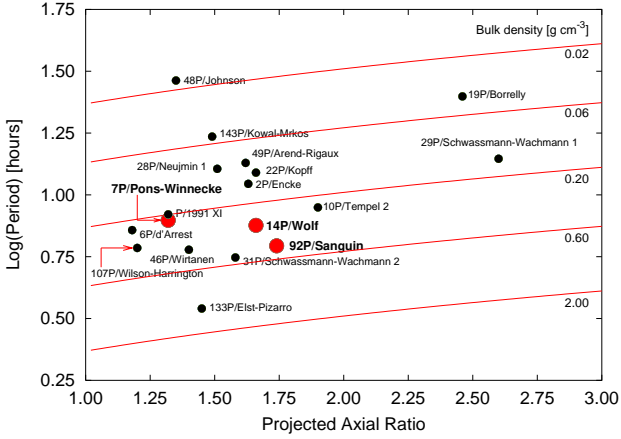}
      \caption{Rotation period against projected axial ratio. The results from 7P/PW, 14P/W and 92P/S are compared with all reliable results from similar studies of JFCs, and of the Halley type comet P/1991 XI. The lines show lower limits on density, from equation \ref{densityeqn}, of 0.02, 0.06, 0.20, 0.60 and 2.00 g cm$^{-3}$. We measure densities of 0.23 g~cm$^{-3}$ for 7P/PW, 0.32 g~cm$^{-3}$ for 14P/W and 0.49 g~cm$^{-3}$ for 92P/S. Data from: 48P/Johnson: Jewitt \& Sheppard \cite{JS2004}; 133P/Elst-Pizarro: Hsieh et al. \cite{H2004}; all others: Lowry \& Weissman \cite{LW2003} and references therein.
              }
         \label{densityplot}
   \end{figure}
%

Figure \ref{densityplot} shows the rotation periods and projected axial ratios for all JFCs (along with Halley-type comet P/1991 XI (Levy), and 29P/Schwassmann-Wachmann 1, which is dynamically a JFC but can also be classed as a Centaur (Lamy et al. \cite{lamy})) that have reliable measurements of these properties. By combining the rotation period and axial ratio, we can put limits on the bulk density of the nucleus. Assuming that it is not rotating faster than the speed which would cause it to break up due to centrifugal forces, the minimum density of the nucleus $D_{\mathrm{N}}$ (g cm$^{-3}$) can be found from:
\begin{equation}\label{densityeqn}
D_{\mathrm{N}} \ge \frac{10.9\mathrm{h}}{P^2_\mathrm{rot}} \frac{a}{b}
\end{equation}
where $P_{\mathrm{rot}}$ is in hours (Pravec \& Harris \cite{PH00}). This assumes that the nucleus has negligible tensile strength; an assumption which is supported by both the break-up of comet Shoemaker-Levy 9 under the gravitational influence of Jupiter (Asphaug \& Benz \cite{AB1996}) and, indirectly, by theoretical models of cometary formation (Donn \cite{D1990}).

Generally, those comets measured so far have been found to have small projected axial ratios, implying that they are not very extended bodies. Although these projected axial ratios only put minimum limits on the actual $a/b$, we expect the rotation axes of observed comets to be randomly orientated. Even with only 18 data points in Fig. \ref{densityplot} we suspect the clustering towards low $a/b$ indicates a real trend towards small elongations. All three of the comets presented here fit into the low axial ratio and short period end of the distribution, with minimum densities of 0.22 $\pm$ 0.08 g cm$^{-3}$ for 7P/PW, 0.32 $\pm$ 0.02 g cm$^{-3}$ for 14P/W and 0.49 $\pm$ 0.06 g cm$^{-3}$ for 92P/S. Lowry \& Weissman (\cite{LW2003}) suggest that there may be a cut off of densities at $D_{\mathrm{N}} \approx 0.6$ g cm$^{-3}$ for cometary nuclei, as is evident in asteroids at $\sim$ 3 g cm$^{-3}$ (Pravec et al. \cite{P2003}). Our data agrees with this hypothesis, however the recent measurement of $D_{\mathrm{N}} \ge 1.3$ g cm$^{-3}$ for 133P/Elst-Pizarro (calculated from the results of Hsieh et al. \cite{H2004}) places it above this limit. We note though that 133P is unusual; thought to be an comet-asteroid transition object, it has asteroid like dynamical properties but has been occasionally reported to show a dust trail (Hsieh et al. \cite{H2004}). Further data is required before it can be decided whether there is a cut off in nuclear densities, or this is simply an observational effect. 

   \begin{figure}
   \centering
   \includegraphics[angle=-90,width=0.47\textwidth]{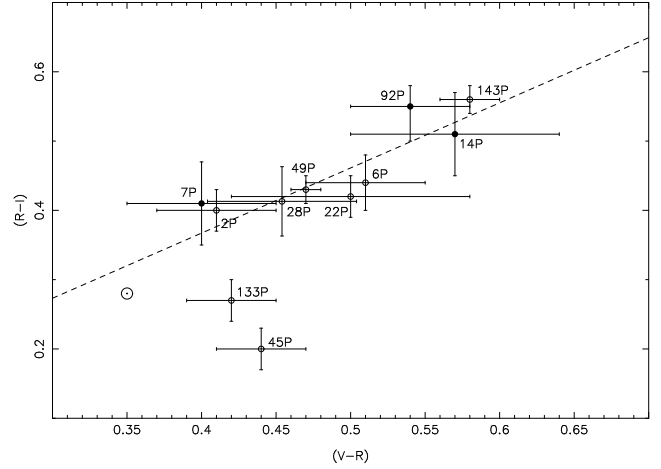}
      \caption{$(R-I)$ against $(V-R)$ for all JFCs with known colours. Filled circles -- this work; open circles -- comets with colours previously determined using the same multi-filter photometry method used here. With the exception of comets 45P/Honda-Mrkos-Pajdusakova and 133P/Elst-Pizarro, all comets are found to be redder than the sun (denoted by symbol $\odot$) in both $(V-R)$ and $(R-I)$. The trend-line for these comets is described in section \ref{discussion} and also here as a dashed line. Data from: 2P/Encke: Jewitt \cite{J2002}; 6P/d'Arrest: Jewitt \cite{J2002}, Lowry \& Weissman \cite{LW2003}; 22P/Kopff: Lamy et al. \cite{La2002}; 28P/Neujmin 1: Delahodde et al. \cite{d2001}; 45P/Honda-Mrkos-Pajdusakova: Lamy et al. \cite{La1999}; 49P/Arend-Rigaux: Millis et al. \cite{M1988}, Lowry et al. \cite{lo2003}; 133P/Elst-Pizarro: Hsieh et al. \cite{H2004}; 143P/Kowal-Mrkos: Jewitt et al. \cite{J2003}. Where more than one reference is given, weighted averages of the values were taken.
              }
         \label{colourplot}
   \end{figure}
%

Figure \ref{colourplot} shows all the available broad-band colours $(R-I)$ against $(V-R)$ for JFCs. Colours of cometary nuclei tell us about surface properties (Davidsson \& Skorov \cite{DS2002}) and allow us to compare comets with other solar system bodies. For example, while short period cometary nuclei are found to be redder than the Sun, they are less red than Kuiper-belt objects, the population from which they are thought to originate. For nuclei, the mean $\overline{(V-R)}_{\mathrm{nuc}} = 0.45 \pm 0.10$ ($N$ = 20), including the results presented here and those collated by Lamy et al. (\cite{lamy}) for other comets. This compares with $\overline{(V-R)}_{\mathrm{KBO}} = 0.60 \pm 0.07$ ($N$ = 71) result for the KBO/TNO population found by combining data from Peixinho et al. (\cite{P2004}) and Jewitt \& Luu (\cite{JL2001}).

Comets 7P/PW, 14P/W and 92P/S fit in amongst the small current sample, with 7P/PW towards the blue end and 14P/W one of the most red objects. Although 7P/PW has the bluest $(V-R)$ in Fig. \ref{colourplot}, the range in $(V-R)$ extends down to slightly bluer than solar, with the comet-asteroid transition object 107P/Wilson-Harrington having $(V-R) = 0.20\pm0.04$ (Lowry \& Weissman \cite{LW2003}). Figure \ref{colourplot} suggests a trend of increasing $(R-I)$ with increasing $(V-R)$, implying that the albedo of these nuclei continues to rise through the $V$, $R$ and $I$ bands. The exceptions to this trend are comets 133P/Elst-Pizarro and 45P/Honda-Mrkos-Pajdusakova. They have $(R-I) < (V-R)$, which implies a downturn of the spectra in the $I$-band similar to silicate dominated asteroids. A best fit straight line to the observed trend:
\begin{equation}
(R-I) =m (V-R) + c \left\{ \begin{array}{l}
	m = 0.94 \pm 0.17\\
	c = 0.01 \pm 0.01
	\end{array} \right.
\end{equation}
also fits the known data for KBOs (Peixinho et al. \cite{P2004}; and Jewitt \& Luu \cite{JL2001}), although there is a large scatter in the KBO results. Again, further data is required to populate this plot and reveal whether this trend is real, or an observational effect. 


\section{Summary}\label{summary}

We present results from $VRI$ photometry of comets 7P/Pons-Winnecke, 14P/Wolf and 92P/Sanguin, obtained using the 3.6m NTT at La Silla in January 2004. We use these data to determine the bulk physical properties of their nuclei. We find:

\begin{enumerate}
\item All three comets were inactive at the time of observation, with upper limits on any unresolved coma placed at $\le 7\%$ of the observed flux for 7P/PW, $\le 5\%$ for 14P/W and $\le 9\%$ for 92P/S.

\item All three comets show clear periodic brightness variations, which we assume to be due to rotation of non-spherical nuclei. Reliable rotation periods were determined at $7.53 \pm 0.10$ and $6.22 \pm 0.05$ hours for 14P/W and 92P/S respectively. There is some ambiguity in the determined rotation period for 7P/PW, which we find to be $6.8 \le P_{\mathrm{rot}} \le 9.5$ hours. Further data is required to constrain this period.

\item We assume the nuclei to have roughly ellipsoidal shapes. This assumption is supported by repeated colour measurements which show no variation, implying that brightness changes are due to geometric considerations, and not due to large scale albedo features. From the measured brightness variation observed, we find minimum axial ratios $a/b$ of $1.3 \pm 0.1$ for 7P/PW, $1.7 \pm 0.1$ for 14P/W and $1.7 \pm 0.1$ for 92P/S.

\item We measure mean effective radii of $2.24 \pm 0.02$ km for 7P/PW, $3.16 \pm 0.01$ km for 14P/W and $2.08 \pm 0.01$ km for 92P/S, assuming an albedo of 0.04 and $\beta$ = 0.035 mag.~deg$^{-1}$. These agree with previous measurements when the measured elongation of these bodies is taken into account.

\item The measured rotation periods and minimum elongations suggest very low minimum bulk densities, in agreement with the rubble-pile model of cometary nuclei. We measure densities of $0.23 \pm 0.08$ g~cm$^{-3}$ for 7P/PW, $0.32 \pm 0.02$ g~cm$^{-3}$ for 14P/W and $0.49 \pm 0.06$ g~cm$^{-3}$ for 92P/S.

\item We measure colour indices of $(V-R) = 0.40 \pm 0.05$ and $(R-I) = 0.41 \pm 0.06$ for 7P/PW, $(V-R) = 0.57 \pm 0.07$ and $(R-I) = 0.51 \pm 0.06$ for 14P/W, and $(V-R) = 0.54 \pm 0.04$ and $(R-I) = 0.54 \pm 0.04$ for 92P/S. These fall in a similar range to previously observed nuclei, being redder than the Sun but bluer than the mean colours found for KBOs. This suggests that cometary activity or other mechanisms have substantially altered the surfaces of these objects since they entered the inner solar system.

\end{enumerate}

\begin{acknowledgements}
We would like to thank O.~Hainaut and the staff at La Silla for their help in these observations, and the referee J.~Licandro for helpful comments. IRAF is distributed by the National Optical Astronomy Observatories, which is operated by the Association of Universities for Research in Astronomy, Inc. (AURA) under cooperative agreement with the National Science Foundation.
\end{acknowledgements}

\clearpage
\setcounter{table}{1}

\begin{table}
\caption{Apparent $R$-band magnitudes for comet 7P/PW.}             
\label{7Ptable}      
\centering                         
\begin{tabular}{c c c c}        
\hline\hline                 
Date & MJD (53020+) & Airmass & $m_R$ \\    
\hline                        
20/01/2004 & 5.0420 & 1.235 & 22.530$\pm$0.096 \\
& 5.0438 & 1.238 & 22.594$\pm$0.099 \\
& 5.0455 & 1.242 & 22.486$\pm$0.118 \\
& 5.0473 & 1.247 & 22.670$\pm$0.108 \\
& 5.0491 & 1.251 & 22.737$\pm$0.113 \\
& 5.0508 & 1.255 & 22.364$\pm$0.112 \\
& 5.0526 & 1.260 & 22.436$\pm$0.116 \\
& 5.0772 & 1.348 & 22.629$\pm$0.095 \\
& 5.0790 & 1.357 & 22.591$\pm$0.096 \\
& 5.0808 & 1.365 & 22.646$\pm$0.099 \\
& 5.0826 & 1.374 & 22.501$\pm$0.089 \\
& 5.0844 & 1.383 & 22.494$\pm$0.089 \\
& 5.0862 & 1.393 & 22.677$\pm$0.103 \\
& 5.0879 & 1.402 & 22.505$\pm$0.090 \\
& 5.1270 & 1.718 & 22.623$\pm$0.101 \\
& 5.1287 & 1.738 & 22.507$\pm$0.095 \\
& 5.1304 & 1.759 & 22.355$\pm$0.104 \\
& 5.1322 & 1.782 & 22.189$\pm$0.096 \\
& 5.1340 & 1.804 & 22.148$\pm$0.092 \\
& 5.1358 & 1.828 & 22.295$\pm$0.103 \\
& 5.1375 & 1.853 & 22.271$\pm$0.104 \\
& 5.1399 & 1.886 & 22.412$\pm$0.119 \\
& 5.1417 & 1.913 & 22.371$\pm$0.134 \\
& 5.1434 & 1.941 & 22.355$\pm$0.133 \\
& 5.1452 & 1.970 & 22.327$\pm$0.124 \\
& 5.1469 & 1.999 & 22.200$\pm$0.099 \\
& 5.1487 & 2.030 & 22.265$\pm$0.106 \\
& 5.1505 & 2.062 & 22.332$\pm$0.114 \\
21/01/2004 & 6.1021 & 1.513 & 22.633$\pm$0.109 \\
& 6.1038 & 1.527 & 22.288$\pm$0.087 \\
& 6.1055 & 1.540 & 22.354$\pm$0.089 \\
& 6.1072 & 1.555 & 22.389$\pm$0.093 \\
& 6.1090 & 1.569 & 22.509$\pm$0.099 \\
& 6.1108 & 1.585 & 22.389$\pm$0.092 \\
& 6.1125 & 1.601 & 22.519$\pm$0.102 \\
& 6.1379 & 1.901 & 22.409$\pm$0.098 \\
& 6.1397 & 1.928 & 22.319$\pm$0.091 \\
& 6.1415 & 1.957 & 22.467$\pm$0.104 \\
& 6.1432 & 1.986 & 22.344$\pm$0.102 \\
& 6.1450 & 2.017 & 22.226$\pm$0.093 \\
& 6.1468 & 2.049 & 22.480$\pm$0.104 \\
& 6.1486 & 2.082 & 22.445$\pm$0.102 \\
\hline                                   
\end{tabular}
\end{table}

\begin{table}
\caption{Apparent $R$-band magnitudes for comet 14P/W.}             
\label{14Ptable}      
\centering                         
\begin{tabular}{c c c c}        
\hline\hline                 
Date & MJD (53020+) & Airmass & $m_R$ \\    
\hline                        
20/01/2004& 5.1734 & 1.817 & 22.818$\pm$0.120 \\
& 5.1763 & 1.774 & 22.635$\pm$0.105 \\
& 5.1792 & 1.734 & 22.704$\pm$0.104 \\
& 5.1821 & 1.696 & 22.665$\pm$0.104 \\
& 5.1850 & 1.660 & 22.604$\pm$0.099 \\
& 5.2244 & 1.318 & 22.419$\pm$0.069 \\
& 5.2273 & 1.300 & 22.333$\pm$0.066 \\
& 5.2302 & 1.284 & 22.395$\pm$0.067 \\
& 5.2331 & 1.268 & 22.367$\pm$0.072 \\
& 5.2360 & 1.253 & 22.316$\pm$0.065 \\
& 5.2542 & 1.173 & 22.212$\pm$0.063 \\
& 5.2571 & 1.163 & 22.197$\pm$0.063 \\
& 5.2600 & 1.152 & 22.252$\pm$0.064 \\
& 5.2629 & 1.143 & 22.208$\pm$0.061 \\
& 5.2659 & 1.133 & 22.204$\pm$0.061 \\
& 5.2853 & 1.083 & 22.213$\pm$0.060 \\
& 5.2882 & 1.077 & 22.266$\pm$0.059 \\
& 5.2912 & 1.071 & 22.187$\pm$0.060 \\
& 5.2941 & 1.066 & 22.197$\pm$0.059 \\
& 5.2970 & 1.061 & 22.207$\pm$0.060 \\
& 5.3150 & 1.039 & 22.329$\pm$0.060 \\
& 5.3179 & 1.036 & 22.273$\pm$0.060 \\
& 5.3208 & 1.034 & 22.333$\pm$0.062 \\
& 5.3237 & 1.032 & 22.389$\pm$0.061 \\
& 5.3266 & 1.030 & 22.316$\pm$0.062 \\
& 5.3299 & 1.029 & 22.398$\pm$0.062 \\
& 5.3328 & 1.028 & 22.410$\pm$0.062 \\
& 5.3361 & 1.027 & 22.540$\pm$0.065 \\
& 5.3390 & 1.027 & 22.507$\pm$0.066 \\
21/01/2004& 6.1868 & 1.606 & 22.193$\pm$0.059 \\
& 6.1897 & 1.575 & 22.239$\pm$0.054 \\
& 6.1973 & 1.500 & 22.113$\pm$0.048 \\
& 6.2002 & 1.474 & 22.109$\pm$0.053 \\
& 6.2277 & 1.282 & 22.098$\pm$0.052 \\
& 6.2306 & 1.266 & 22.189$\pm$0.053 \\
& 6.2335 & 1.251 & 22.103$\pm$0.050 \\
& 6.2363 & 1.237 & 22.094$\pm$0.050 \\
& 6.2392 & 1.224 & 22.165$\pm$0.052 \\
& 6.2586 & 1.147 & 22.279$\pm$0.052 \\
& 6.2614 & 1.138 & 22.243$\pm$0.048 \\
& 6.2643 & 1.129 & 22.409$\pm$0.053 \\
& 6.2672 & 1.121 & 22.299$\pm$0.049 \\
& 6.2701 & 1.113 & 22.419$\pm$0.063 \\
& 6.2885 & 1.071 & 22.407$\pm$0.052 \\
& 6.2913 & 1.066 & 22.419$\pm$0.053 \\
& 6.2942 & 1.061 & 22.265$\pm$0.048 \\
& 6.2971 & 1.057 & 22.418$\pm$0.052 \\
& 6.3000 & 1.053 & 22.349$\pm$0.049 \\
& 6.3080 & 1.043 & 22.334$\pm$0.049 \\
& 6.3108 & 1.040 & 22.346$\pm$0.047 \\
& 6.3137 & 1.037 & 22.421$\pm$0.054 \\
& 6.3166 & 1.035 & 22.385$\pm$0.048 \\
& 6.3195 & 1.033 & 22.285$\pm$0.045 \\
& 6.3296 & 1.028 & 22.245$\pm$0.045 \\
& 6.3325 & 1.027 & 22.267$\pm$0.046 \\
& 6.3354 & 1.027 & 22.213$\pm$0.044 \\
& 6.3383 & 1.027 & 22.255$\pm$0.046 \\
& 6.3411 & 1.027 & 22.206$\pm$0.044 \\
\hline                                   
\end{tabular}
\end{table}

\begin{table*}[p]
\begin{minipage}{2\columnwidth}
\caption{Apparent $R$-band magnitudes for comet 92P/S.}             
\label{92Ptable}
\centering
\begin{tabular}{c c c c c c c c}        
\hline\hline                 
Date & MJD (53020+) & Airmass & $m_R$ & Date & MJD (53020+) & Airmass & $m_R$\\    
\hline           
20/01/2004& 5.0629 & 1.304 & 21.924$\pm$0.079 & & 5.3063 & 2.237 & 22.091$\pm$0.117\\
& 5.0642 & 1.298 & 21.868$\pm$0.075 & & 5.3075 & 2.267 & 22.422$\pm$0.174\\
& 5.0654 & 1.293 & 21.892$\pm$0.080 & & 5.3088 & 2.298 & 22.288$\pm$0.140\\
& 5.0667 & 1.288 & 21.871$\pm$0.065 & & 5.3100 & 2.329 & 22.081$\pm$0.126\\
& 5.0679 & 1.283 & 21.771$\pm$0.057 & & 5.3113 & 2.362 & 22.447$\pm$0.181\\
& 5.0692 & 1.278 & 21.860$\pm$0.064 & & 5.3125 & 2.395 & 22.556$\pm$0.199\\
& 5.0704 & 1.273 & 21.813$\pm$0.060 & 21/01/2004& 6.1241 & 1.151 & 21.631$\pm$0.075 \\
& 5.0717 & 1.268 & 21.856$\pm$0.063 & & 6.1253 & 1.151 & 21.570$\pm$0.071\\
& 5.0730 & 1.263 & 21.796$\pm$0.059 & & 6.1265 & 1.150 & 21.525$\pm$0.068\\
& 5.0742 & 1.258 & 21.752$\pm$0.070 & & 6.1277 & 1.149 & 21.705$\pm$0.092\\
& 5.0991 & 1.189 & 21.758$\pm$0.056 & & 6.1289 & 1.149 & 21.520$\pm$0.066\\
& 5.1004 & 1.186 & 21.772$\pm$0.056 & & 6.1302 & 1.148 & 21.707$\pm$0.092\\
& 5.1017 & 1.184 & 21.827$\pm$0.058 & & 6.1315 & 1.148 & 21.581$\pm$0.081\\
& 5.1029 & 1.181 & 21.777$\pm$0.056 & & 6.1327 & 1.147 & 21.671$\pm$0.088\\
& 5.1041 & 1.179 & 21.816$\pm$0.057 & & 6.1340 & 1.147 & 21.629$\pm$0.085\\
& 5.1054 & 1.177 & 21.785$\pm$0.057 & & 6.1352 & 1.147 & 21.733$\pm$0.094\\
& 5.1067 & 1.175 & 21.879$\pm$0.062 & & 6.1600 & 1.157 & 21.734$\pm$0.058\\
& 5.1079 & 1.173 & 21.913$\pm$0.078 & & 6.1624 & 1.159 & 21.827$\pm$0.063\\
& 5.1092 & 1.171 & 21.797$\pm$0.057 & & 6.1636 & 1.161 & 21.881$\pm$0.068\\
& 5.1105 & 1.169 & 21.893$\pm$0.075 & & 6.1648 & 1.162 & 21.749$\pm$0.070\\
& 5.1610 & 1.154 & 22.343$\pm$0.096 & & 6.1661 & 1.164 & 21.975$\pm$0.071\\
& 5.1622 & 1.155 & 22.171$\pm$0.076 & & 6.1672 & 1.165 & 21.840$\pm$0.064\\
& 5.1634 & 1.156 & 22.163$\pm$0.077 & & 6.1685 & 1.167 & 21.872$\pm$0.064\\
& 5.1647 & 1.158 & 22.168$\pm$0.077 & & 6.1697 & 1.168 & 21.983$\pm$0.070\\
& 5.1659 & 1.159 & 22.442$\pm$0.095 & & 6.1709 & 1.170 & 21.930$\pm$0.069\\
& 5.1672 & 1.160 & 22.242$\pm$0.080 & & 6.2048 & 1.253 & 22.041$\pm$0.094\\
& 5.1684 & 1.162 & 22.249$\pm$0.083 & & 6.2060 & 1.257 & 22.023$\pm$0.096\\
& 5.1696 & 1.163 & 22.371$\pm$0.096 & & 6.2072 & 1.262 & 21.936$\pm$0.090\\
& 5.1709 & 1.165 & 22.403$\pm$0.101 & & 6.2084 & 1.266 & 21.896$\pm$0.095\\
& 5.2104 & 1.261 & 21.911$\pm$0.068 & & 6.2097 & 1.271 & 22.070$\pm$0.098\\
& 5.2116 & 1.265 & 22.028$\pm$0.077 & & 6.2109 & 1.275 & 22.029$\pm$0.111\\
& 5.2128 & 1.270 & 21.989$\pm$0.083 & & 6.2121 & 1.280 & 21.926$\pm$0.099\\
& 5.2141 & 1.275 & 22.011$\pm$0.077 & & 6.2133 & 1.285 & 21.806$\pm$0.075\\
& 5.2154 & 1.280 & 21.841$\pm$0.075 & & 6.2145 & 1.290 & 21.868$\pm$0.091\\
& 5.2166 & 1.285 & 21.892$\pm$0.077 & & 6.2157 & 1.295 & 21.847$\pm$0.091\\
& 5.2179 & 1.290 & 21.881$\pm$0.065 & & 6.2451 & 1.461 & 21.611$\pm$0.076\\
& 5.2192 & 1.296 & 21.818$\pm$0.064 & & 6.2463 & 1.470 & 21.572$\pm$0.072\\
& 5.2205 & 1.301 & 21.966$\pm$0.072 & & 6.2475 & 1.479 & 21.738$\pm$0.076\\
& 5.2217 & 1.307 & 21.920$\pm$0.069 & & 6.2487 & 1.488 & 21.600$\pm$0.079\\
& 5.2403 & 1.406 & 21.860$\pm$0.066 & & 6.2499 & 1.498 & 21.645$\pm$0.083\\
& 5.2416 & 1.414 & 21.873$\pm$0.068 & & 6.2511 & 1.507 & 21.761$\pm$0.094\\
& 5.2428 & 1.423 & 21.908$\pm$0.070 & & 6.2523 & 1.517 & 21.556$\pm$0.079\\
& 5.2441 & 1.431 & 21.938$\pm$0.072 & & 6.2535 & 1.527 & 21.611$\pm$0.082\\
& 5.2453 & 1.440 & 22.006$\pm$0.077 & & 6.2547 & 1.538 & 21.727$\pm$0.108\\
& 5.2466 & 1.448 & 21.909$\pm$0.081 & & 6.2560 & 1.549 & 21.673$\pm$0.091\\
& 5.2478 & 1.457 & 22.200$\pm$0.107 & & 6.2742 & 1.741 & 21.820$\pm$0.105\\
& 5.2491 & 1.466 & 21.876$\pm$0.071 & & 6.2754 & 1.757 & 21.850$\pm$0.107\\
& 5.2503 & 1.476 & 21.997$\pm$0.088 & & 6.2766 & 1.772 & 21.927$\pm$0.114\\
& 5.2516 & 1.485 & 21.958$\pm$0.073 & & 6.2778 & 1.788 & 21.840$\pm$0.109\\
& 5.2701 & 1.655 & 22.107$\pm$0.099 & & 6.2791 & 1.805 & 21.855$\pm$0.106\\
& 5.2713 & 1.669 & 22.029$\pm$0.093 & & 6.2803 & 1.822 & 21.777$\pm$0.098\\
& 5.2726 & 1.683 & 22.103$\pm$0.095 & & 6.2815 & 1.839 & 21.870$\pm$0.109\\
& 5.3012 & 2.125 & 22.120$\pm$0.129 & & 6.2827 & 1.857 & 21.804$\pm$0.102\\	
& 5.3025 & 2.152 & 22.522$\pm$0.193 & & 6.2839 & 1.874 & 21.832$\pm$0.108\\	
& 5.3038 & 2.180 & 22.457$\pm$0.172 & & 6.2851 & 1.893 & 21.712$\pm$0.094\\	
& 5.3050 & 2.208 & 22.408$\pm$0.168 \\
\hline                                   
\end{tabular}
\end{minipage}
\end{table*}


\begin{thebibliography}{}

  \bibitem[1996]{AB1996} Asphaug, E., Benz, W., 1996,
  		Icarus, 121, 225
  \bibitem[2002]{DS2002} Davidsson, B., Skorov, Y., 2002,
  		Icarus, 156, 223
  \bibitem[1999]{IRAF3} Davis, L., 1999,
      	in ASP Conf. Ser. 189, Precision CCD Photometry,
	ed. Craine, E., Crawford, D., \& Tucker, R.,
	35
  \bibitem[2001]{d2001} Delahodde, C., Meech, K., Hainaut, O., Dotto, E., 2001,
      	A\&A, 376, 672
  \bibitem[1990]{D1990} Donn, B., 1990,
      	A\&A, 235, 441
  \bibitem[2004]{d2004} Duxbury, T., Newburn, R., Brownlee, D., 2004
      	\jgr 109, E12S02
  \bibitem[2003]{NTT-illum} Hainaut, O., 2003,
      ESO La Silla SciOps Technical Report: LSO-TRE-ESO-40200-1061/1.0,
      http://www.ls.eso.org/lasilla/sciops/doc/LSO-TRE-ESO\_40200-1061\_susiBaffle/LSO-TRE-ESO\_40200-1061\_susiBaffle.html
  \bibitem[1989]{ap-FWHM} Howell, S., 1989,
  		PASP, 101, 616
  \bibitem[2004]{H2004} Hsieh, H., Jewitt, D., Fern\'andez, Y., 2004,
      	AJ, 127, 2997
  \bibitem[1997]{if1997} Ip, W.-H., Fern\'andez, J., 1997,
      	A\&A, 324, 778
  \bibitem[1984]{JD84} Jewitt, D., Danielson, G., 1984,
  		Icarus, 60, 435
  \bibitem[2001]{JL2001} Jewitt, D., Luu, J., 2001,
      	AJ, 122, 2099
  \bibitem[2002]{J2002} Jewitt, D., 2002,
      	AJ, 123, 1039
  \bibitem[2003]{J2003} Jewitt, D., Sheppard, S., Fern\'andez, Y., 2003,
      	AJ, 125, 3366
  \bibitem[2004]{JS2004} Jewitt, D., Sheppard, S., 2004,
      	AJ, 127, 1784
  \bibitem[1999]{La1999} Lamy, P., Toth, I., A'Hearn, M., Weaver, H., 1999,
  		Icarus, 140, 424
  \bibitem[2002]{La2002} Lamy, P., Toth, I., Jorda, L., et al., 2002,
  		Icarus, 156, 442
  \bibitem[2005]{lamy} Lamy, P., Toth, I., Fern\'andez, Y., Weaver, H., 2005,
        in Comets II,
      ed. Festou, M., Keller, H., \& Weaver, H.
      (University of Arizona Press) in press.
  \bibitem[1992]{landolt} Landolt, A., 1992,
      	AJ, 104, 340
  \bibitem[2000b]{Lic2000} Licandro, J., Serra-Ricart, M., Oscoz, A., Casas, R., Osip, D., 2000b,
  		AJ, 119, 3133
  \bibitem[2000]{li2000} Licandro, J., Tancredi, G., Lindgren, M., Rickman, H., Gil-Hutton, R., 2000,
  		Icarus, 147, 161
  \bibitem[1976]{Lomb} Lomb, N., 1976,
  		A\&SS, 39, 447
  \bibitem[2001]{LF2001} Lowry, S., Fitzsimmons, A., 2001,
      	A\&A, 365, 204
  \bibitem[2003]{lo2003} Lowry, S., Fitzsimmons, A., Collander-Brown, S., 2003,
      	A\&A, 397, 329
  \bibitem[2003]{LW2003} Lowry, S., Weissman, P., 2003,
  		Icarus, 164, 492
  \bibitem[2005]{LF2005} Lowry, S., Fitzsimmons, A., 2005,
      	MNRAS, 358, 641
  \bibitem[2004]{M2004} Meech, K. Hainaut, O. Marsden, B., 2004,
  		Icarus, 170, 463
  \bibitem[1988]{M1988} Millis, R., A'Hearn, M., Campins, H., 1988,
  		APJ, 324, 1194
  \bibitem[2004]{o2004} Oberst, J., Giese, B., Howington-Kraus, E., et al., 2004,
  		Icarus, 167, 70
  \bibitem[2004]{P2004} Peixinho, N., Boehnhardt, H., Belskaya, I., et al., 2004,
  		Icarus, 170, 153
  \bibitem[2000]{PH00} Pravec, P., Harris, A., 2000,
  		Icarus, 148, 12
  \bibitem[2003]{P2003} Pravec, P., Harris, A., Michalowski, T., 2003,
        in Asteroids III,
      ed. Bottke, W., Cellino, A., Paolicchi, P., \& Binzel, R.
      (University of Arizona Press) page 113
  \bibitem[1916]{R16} Russel, H., 1916,
  		APJ, 43, 173
  \bibitem[1991]{sa1991} Stooke, P., Abergel, A., 1991,
      	A\&A, 248, 656
	
  \bibitem[1986]{IRAF1}	Tody, D. 1986, 
	In Proc. SPIE Instrumentation in Astronomy VI.
	ed. Crawford, D. 627, 733

  \bibitem[1993]{IRAF2}	Tody, D. 1993, 
	In Astronomical Data Analysis Software and Systems II, A.S.P. Conf. Ser. 52
	ed. Hanisch, R., Brissenden, R., \& Barnes, J., 173


\end{thebibliography}
\end{document}